# Plasmon dynamics in graphene

Authors: Suheng Xu[1+*], Birui Yang[1+], Nishchhal Verma[1], Rocco A. Vitalone[1], Brian Vermilyea[2], Miguel Sánchez Sánchez[1,3], Julian Ingham[1], Ran Jing[4,5], Yinming Shao[1,11], Tobias Stauber[3], Angel Rubio[6,7,8], Milan Delor[9], Mengkun Liu[4,10], Michael M. Fogler[2], Cory R. Dean[1], Andrew Millis[1,7], Raquel Queiroz[1], D. N. Basov[1*]
Affiliations:
[1]*Department of Physics, Columbia University, New York, New York 10027, USA*
[2]*Department of Physics, University of California at San Diego, La Jolla, CA 92093-0319, USA*
[3]*Instituto de Ciencia de Materiales de Madrid ICMM-CSIC, Madrid, Spain*
[4]*Department of Physics and Astronomy, Stony Brook University, Stony Brook, New York 11794, USA*
[5]*Condensed Matter Physics and Materials Science Department, Brookhaven National Laboratory, Upton, New York 11973, USA*
[6]*Theory Department, Max Planck Institute for Structure and Dynamics of Matter and Center of Free-Electron Laser Science, 22761 Hamburg, Germany*
[7]*Center for Computational Quantum Physics, The Flatiron Institute, 162 5th Avenue, New York, New York 10010, USA*
[8]*Nano-Bio Spectroscopy Group, Universidad del País Vasco UPV/EHU, San Sebastián 20018, Spain*
[9]*Department of Chemistry, Columbia University, New York, NY, 10027, USA*
[10]*National Synchrotron Light Source II, Brookhaven National Laboratory, Upton, New York 11973, USA*
[11]*Department of Physics, Pennsylvania State University, University Park, PA, 16802, USA*
+*These authors contributed equally to the work*
**Corresponding authors: Suheng Xu: sx2277@columbia.edu, D. N. Basov: db3056@columbia.edu*

**ABSTRACT**. Plasmons are collective oscillations of mobile electrons with dynamics controlled by their charge stiffness ("Drude weight"). Using terahertz spacetime metrology, we probe plasmon dynamics of mono- and bi-layer graphene. In both systems, the experimentally measured Drude weight systematically exceeds the prediction based on non-interacting electronic system. The relative enhancement increases as the carrier density decreases. We attribute the observed deviation to the interplay of interactions and wave function structure of the Dirac fermions in multi-layer graphene. Our results establish that pseudospin structure of the single-particle electronic wave function can directly influence collective excitations, with implications that extend beyond graphene to a broad class of quantum materials.

## I. INTRODUCTION.

In 1900, Paul Drude introduced a classical model of electronic transport that accurately describes both electrical conductivity and long-wavelength plasmonic oscillations in a wide range of materials [1]. This surprising success is rooted in Galilean invariance, a foundational principle governing transport and collective excitations in solids. In Galilean-invariant systems, although the plasmon frequency is set by Coulomb interaction, its dispersion relation are fully determined by bare single-particle quantities---carrier density and band mass, with no additional renormalization from dynamical electron-electron interactions [2–11].

Galilean invariance arises from the continuous translation invariance of interacting Hamiltonians in free space, and in crystals is broken down to a discrete translational invariance by the presence of the lattice. However, in many materials the energy bands are approximately parabolic in the relevant energy regimes, leading to an emergent Galilean invariant electron fluid characterized by a free carrier density and band mass. Here, we demonstrate that in monolayer and Bernal bilayer graphene, despite the emergent continuous translation invariance, electronic interactions fundamentally modify collective plasmon dynamics, and we attribute this behavior to a momentum dependent orbital structure of the wave function, arising from the discrete translational invariance of the underlying crystal and described in terms of a quantum geometry or pseudospin texture [12,13].

In monolayer graphene (MLG) and Bernal bilayer graphene (BLG) the low energy single particle electronic states are coherent but momentum-dependent superpositions of different orbitals across sublattices or layers, with a momentum-dependent relative phase. For an N-layer graphene system, the conduction-band wavefunction can be written as $|\psi_k\rangle = (|a_k\rangle + |b_k\rangle e^{Ni\phi_k})/\sqrt{2}$, where $\phi_k$ is the azimuthal angle of in-plane momentum $(k_x, k_y)$ [12,14]. Here, $|a_k\rangle$ and $|b_k\rangle$ represent orbitals localized onto the two sublattices in MLG, or the two layers in BLG or the two outer layers in rhombohedral graphene. The orbital content of such a wave function can be represented by a pseudospin: a unit vector $n_k = (\cos(N\phi_k), \sin(N\phi_k), 0$. As shown in Fig.1(a-d), the pseudospin describing the conduction band states winds once and twice as the wave vector winds around the Dirac or quadratic band touching point of the Brillouin zone for MLG and BLG, respectively. This winding implies a quantum geometry that includes a momentum-space vortex structure centered at the Dirac/quadratic band touching point. The quantum geometry transforms non-trivially under a change of reference frame (Fig. 1(h-j), and the winding means that the change cannot be gauged away. These nontrivial transformation properties correspond to a breaking of continuous translation invariance reflected not in the dispersion but in the wave function structure and are responsible for the renormalization of plasmon dispersion predicted in Ref. [5]. The impact of graphene's nontrivial electronic structure on collective electron dynamics is anticipated to be strongest in the vicinity of charge neutrality point where the winding is most rapid. Thus, the experimental studies of the plasmon renormalization carried out in very weakly doped graphene provide a direct pathway for exploring quantum geometric effects in the collective responses.

Here, we use newly developed technique of nano-THz spacetime metrology [15,16] to investigate plasmon dynamics in MLG and BLG. By visualizing the spacetime trajectories (worldlines) of plasmonic waves, we extract plasmon velocities and Drude weights as a function of the carrier density. We find that both MLG and BLG exhibit a density-dependent enhancement of the Drude weight relative to noninteracting theory predictions, with the relative enhancement increasing as the density decreases to the charge neutrality point. The findings are contrary to conventional expectations that in low carrier density systems plasmon velocities are not renormalized by interaction and are also contrary to conventional expectations that the main effect of interactions is to suppress charge motion by increasing carrier mass. Our observed enhancement reveals a non-trivial interplay between electron-electron correlations and quantum geometry and establishes plasmonic worldline metrology as a powerful tool for probing many-body correlations in solids in space and time.

## II. METHODS

We employ scattering-type scanning near-field optical microscope (s-SNOM) [17] to visualize propagating plasmons at THz frequencies far below the diffraction limit [supplemental materials]. This technique accesses large in-plane momenta with a spatial resolution of ~20-50nm [18–22]. As illustrated in Fig. 1e, picosecond THz pulses spanning the range between 0.5-1.5 THz illuminate both the sample and the metalized atomic force microscope (AFM) tip [23–29]. Raster-scanning the sample beneath the tip yields two-dimensional maps of the local THz response of encapsulated graphene devices, including both monolayer and Bernal bilayer graphene. For clarify, Fig. 1f shows representative data acquired on BLG. In both MLG and BLG, the scattered THz amplitude from graphene exceeds that of the SiO2 substrate, confirming the high conductivity under back-gate modulation.

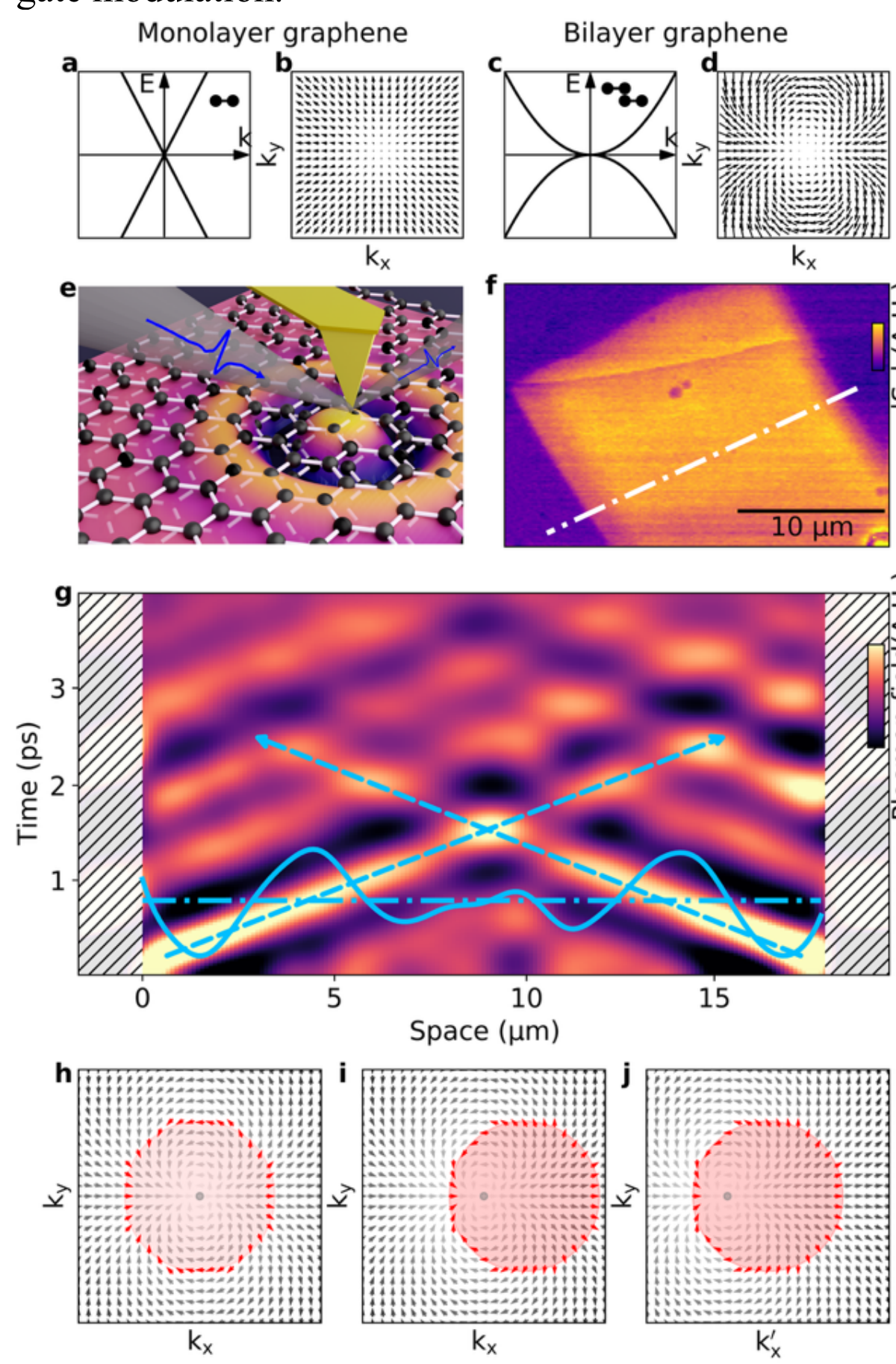

***Figure. 1.*** *Electronic structure of graphene, pseudospin textures and plasmon dynamics . (a) The single-particle electronic band structure of MLG. (b) Pseudospin of conduction band of MLG in momentum space. (c) The single-particle electronic band structure of BLG within a minimal two-band model, neglecting trigonal warping. (d) Pseudospin of conduction band of BLG in momentum space. (e) Schematics of THz near-field measurement on a hBN-encapsulated graphene residing on $SiO_2$/Si substrate. (f) THz near-field image of the BLG sample. The false-color map corresponds to the amplitude of the tip-scattered THz pulses. (g) A spacetime map of plasmon in BLG measured at $V_{bg} - V_{CNP} = 15V$ , T=38K. The shaded regions correspond to the substrate area and the false color map shows the plasmon field strength. (h-j) Schematics of pseudospin texture in BLG in momentum space, illustrating the occupation states at different*

*stages of a plasmon oscillation. The red circle denotes the Fermi surface, and the red arrows indicate the pseudospin orientation of electrons along it. (h) Equilibrium configuration. (i) Under a THz field, the Fermi surface is displaced in momentum space (j) Galilean transformation of (i), corresponding to a reference frame moving with the center of mass. The transformation does not restore the original pseudospin configuration in (h), revealing the absence of Galilean invariance and the coupling between plasmon dynamics and pseudospin texture.*

In all measurements, plasmons are launched via tip-sample near-field coupling. The resulting plasmonic wave packets propagate radially away from the tip, reflect at the graphene edge, and return to the tip where it is detected. To capture this motion, we utilize nano-THz spacetime mapping [15,16,30,31]. This latter technique provides a direct measurement of plasmon velocity, from which the Drude weight of the electron liquid may be determined as we will discuss below. Fig. 1g shows a representative plasmonic spacetime map acquired for BLG at $V_{bg} - V_{CNP} = 15V$, T=38K. Line cuts at fixed time delays (sold lines in Fig 1g) reveal the propagating wave packets. The false-color scale encodes the measured electric field strength of the plasmon wave packet, proportional to the local charge density variation. The horizontal axis corresponds to the tip position along the white dashed line in Fig. 1f, with the shaded region marking the $SiO_2$ substrate beyond the BLG edge. The slope of the worldlines quantifies the plasmon group velocity $v_g$, while the temporal decay of the field amplitude along the worldline determines the plasmon lifetime [supplemental material]. A series of such spacetime maps acquired on both MLG and BLG will be presented and analyzed in the following section.

## III. RESULTS

We mapped the plasmonic worldlines at T=45 K across a range of back-gate voltages. Representative data sets for MLG and BLG at selected $V_{bg}$ are shown in Fig. 2a. With increasing $V_{bg}$, the worldline slopes $dt/dx$ decrease, indicating higher plasmon group velocities. The persistence of plasmon worldlines at CNP is noteworthy. This latter finding enables direct probing of the collective response of the electron liquid in the ultra-dilute regime.

Figures 2b-c display the gate-voltage dependent group velocity extracted from the spacetime measurements. In both systems, $v_g$ decreases toward a minimum value at CNP. At comparable carrier densities, plasmons in MLG propagate faster than those in BLG, reflecting the steeper slope of the electronic Dirac dispersion in MLG. In BLG, the plasmon dispersion further exhibits a pronounced electron-hole asymmetry: plasmons on the electron-dopped side (positive $V_{bg}$) propagate faster than those on the hole-doped side at the same carrier density. This asymmetry arises partly from the intrinsic distinctions between conduction and valance bands in the BLG [32]. In addition, the single back-gate architecture of our devices introduces a nonzero displacement field at nonzero gate voltage, which is enhanced under hole doping due to intrinsic offset doping but is negligible on the electron side [supplemental material].

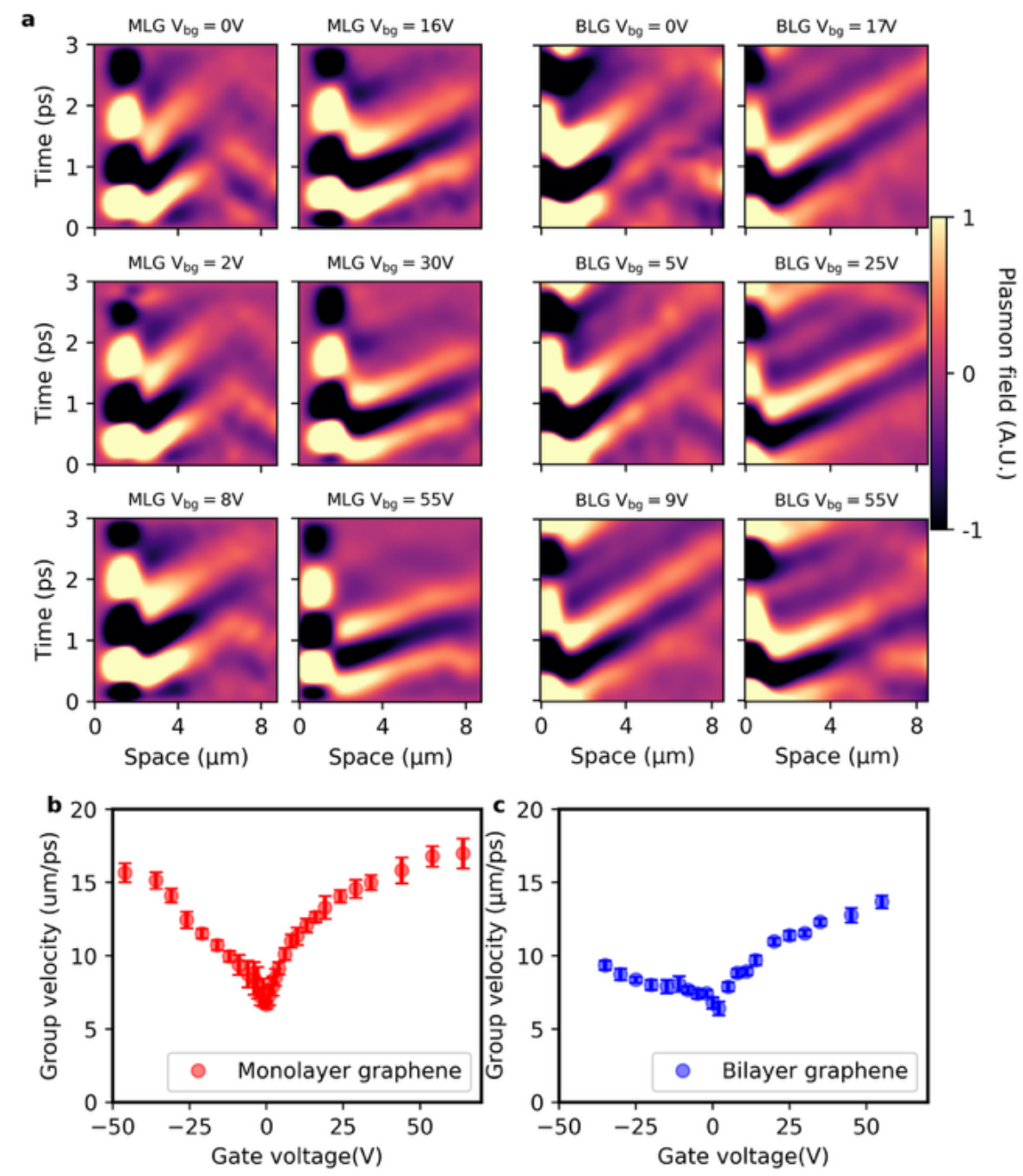

***Figure. 2**. (a) Representative spacetime maps for MLG and BLG collected at 45K for selected gate voltages ($V_{bg} = 0$ corresponds to the charge neutrality point). (b-c) Carrier density dependent group velocity of plasmon in MLG and BLG at T=45K.*

To quantify the electrodynamic response, we evaluate the Drude weight, or the charge stiffness, which characterizes the rigidity of the electron liquid to an applied electric field. The Drude weight is directly extracted from the worldline metrology as $D = \frac{v_g^2 \pi \epsilon \epsilon_0}{d}$. (Supplemental material). Here, d is the distance between the graphene layer and the metallic back gate, and $\epsilon$ is the dielectric constant of the spacer separating them. In a Galilean-invariant system, the Drude weight is given by $D_0 = \frac{e^2 n}{m_0}$, where n is the carrier density and $m_0$ is the effective mass. Thus, one expects a linear dependence of D on n with a zero intercept as n goes to zero.

Figure 3a-b displays the experimentally extracted Drude weights ($D$) as a function of the carrier density. We focus on the electron-doped regime, where displacement-field effects are weaker and the plasmonic features are more pronounced. Although a finite displacement-field is present in a single-gate device under back-gate bias, its effect on the electronic bands

is small and therefore does not affect the conclusions of this work [32] . At doping below $1 - 2 \times 10^{11} cm^{-2}$, as indicated by the shaded region, charge dynamics are dominated by disorder and thermal fluctuations, making an exact determination of the carrier density inaccessible (Supplemental material). For both MLG and BLG, the Drude weight D decreases with decreasing carrier density, but deviates from the naïve density scaling expected in Lorentz/Galilean-invariant systems. For a quantitative comparison it is important to note that in actual materials the single particle band structure deviates slightly from the simple linear/quadratic velocity, in particular because of trigonal warping [14] . To show that the deviations go beyond this band theory effect, in Fig 3(a,b) we compare our data with the results obtained from a non-interacting model using the accepted band theory of the two systems.

The Drude weight renormalization, quantified by the ratio $D/D_0$, is enhanced with decreasing carrier density, as shown in Fig.3(c-d). In MLG, the enhancement is weakly density-dependent and consistent with a logarithmic enhancement (gray dashed line in panel a; black dashed in c); in BLG the enhanced spectral weight scales approximately as $1/\sqrt{n}$.

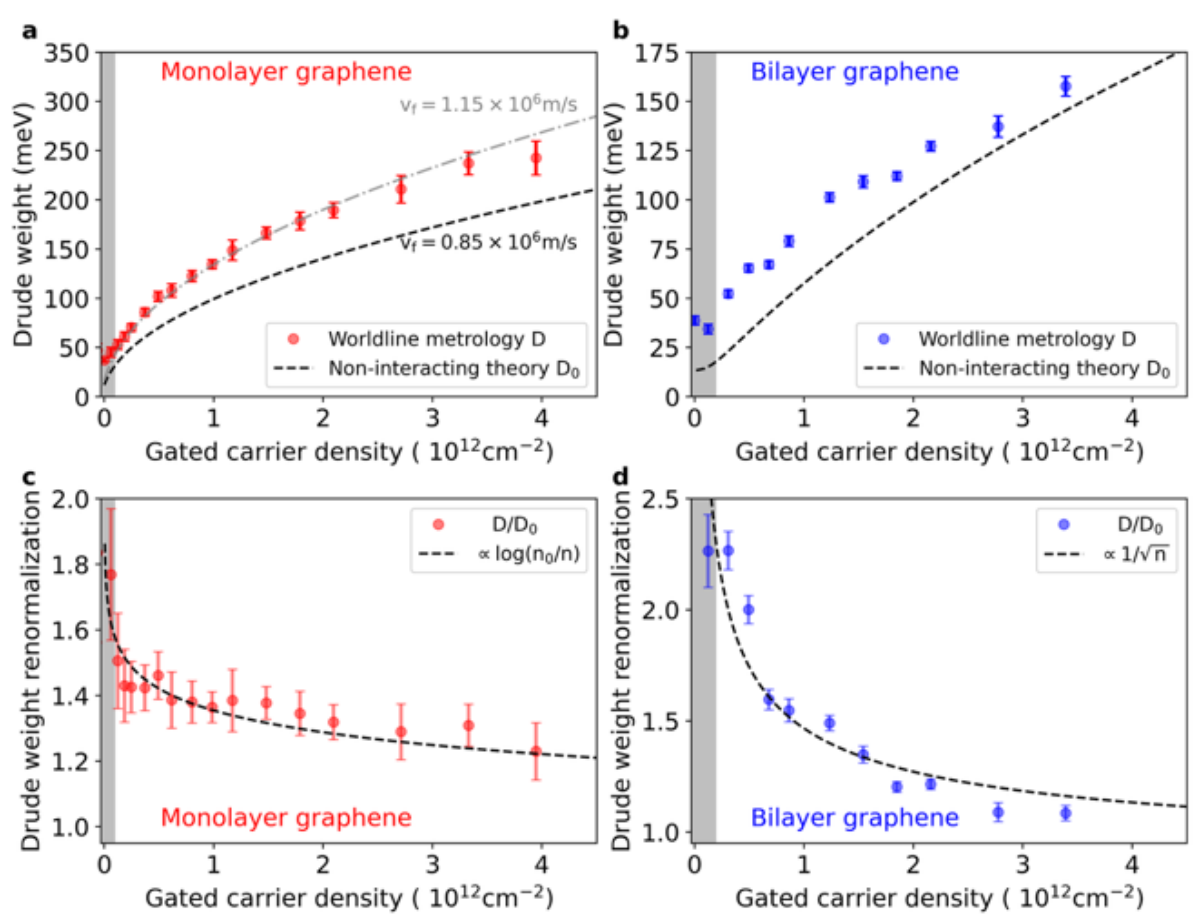

***Figure. 3***. *Plasmonic probe of the Drude weight renormalization. (a) Carrier-density-dependent Drude weight in monolayer graphene. The gray shaded region indicates the disorder-dominated regime. The black dashed line shows the theoretical Drude weight for MLG with linear Dirac dispersion:* $D_0 = \frac{e^2 v_F}{\sqrt{\pi} \hbar^2}\sqrt{n}$, *calculated using the bare Fermi velocity* $v_F = 0.85 \times 10^6 m/s$ [33]. *The gray dashed line corresponds to a Dirac theory with an enhanced fermi velocity of* $v_F = 1.15 \times 10^6 m/s$. *(b) Carrier-density-dependent Drude weight in bilayer graphene. The gray shaded region denotes the thermal regime. The black dashed line represents the theoretical estimated Drude weight calculated using a four-band tight-binding model with best-fit hopping parameters* [34].- *The theoretical estimates in both (a) and (b) assume an electronic temperature T=40 K.(c-d) The ratio between the experimentally measured Drude weight and non-interacting Drude weight in monolayer and bilayer graphene. The dashed traces indicate a phenomenological fit to the carrier density dependent renormalization.*

## IV. The physics of the Drude weight renormalization

We now show that the Drude weight renormalization in graphene revealed by plasmon dynamics (Figs.2,3) can be understood because of the interplay of pseudospin physics and interactions. A plasmon corresponds to oscillations of current and charge. At equilibrium, the Fermi surface is centered at $k = 0$, with a radius set by carrier density (blue circles in Figs. 1h). An electric current corresponds to the momentum-space shift of the Fermi surface (red circle in Figs.1h,i). In conventional 2D electron gases, Galilean invariance ensures that such a shift leaves the internal structure of the wave function unchanged, so plasmon dynamics is governed by the center-of-mass motion and scales as $n/m_0$.

By contrast, in both MLG and BLG, shifting the Fermi surface alters the pseudospin over the entire Fermi surface (Fig. 1i). A simple coordinate transformation (Fig. 1j) cannot restore the original configuration, revealing the coupling between plasmon dynamics and pseudospin textures. Consequently, plasmonic oscillations are inherently coupled to sublattice (MLG) or interlayer (BLG) degrees of freedom. The energy cost associated with these pseudospin rearrangements stiffens the plasmon mode, thereby increasing its propagation velocity [5] . Thus, our THz metrology data establish a direct link between the plasmonic response and the pseudospin texture of the electron wavefunction in the presence of electron-electron interactions.

The coupling between charge and pseudospin dynamics can be understood within perturbation theory [5]. We schematically show these corrections in Fig. 4a. The Drude weight is given by the static, long-wavelength limit of the current-current correlator, quantifying the ability of an applied electric field to generate particle–hole excitations. In the non-interacting limit, it is captured by an intra-band response and depends entirely by the single-particle band dispersion (middle panel in Fig. 4a). Interactions, however, open additional pathways for the creation of particle-hole excitations.

In the first order in perturbation theory, interactions may lead to two distinct types of corrections. The self-energy corrections (left panel in Fig.4a) dress the electronic Greens function, renormalizing the quasiparticle dispersion effectively reshaping the band structure. In systems with a continuous translation invariance, these self-energy corrections are largely cancelled by intra-band interaction-induced vertex corrections [2–11]. However, in the systems studied here, there is an additional Drude-interband mixing [4,35], in which the electric field generates an interband transition and the Coulomb interaction downscatters the virtual particle-hole pair into the conduction band. Unlike self-energy corrections, these vertex corrections are not captured by probes such as Shubnikov-de Haas oscillations or angle-resolved photoemission spectroscopy [2]. Yet, we find that the inter-/intra-band mixing manifests in THz space-time metrology (Figs.2,3). They have a non-trivial doping dependence because they involve the wave function winding, which becomes more rapid as the wavevector approaches the momentum space vortex at the Dirac/band touching point. As a result, Drude weight acquires a genuine new geometric contribution due to the coupling between charge and pseudospin dynamics through the Coulomb interaction.

To further quantify the Drude weight renormalization in MLG and BLG, we compute the leading (first order) interaction corrections within perturbation theory [supplemental material]. Fig. 4b shows the calculated Drude weight renormalization in MLG for different dielectric constant $\epsilon$, representing different Coulomb interaction strength, modeled as $v_q = e^2/2\epsilon\epsilon_0|k - k'|$ with k and k' two relevant electronic wave vectors. In the limit of complete screening ( $\epsilon \to \infty$ ), the renormalization vanishes. At finite $\epsilon$, however, the Drude weight is enhanced, with the effect becoming stronger at lower carrier densities or smaller screening constants. Importantly, as shown in the supplementary material, for MLG even the local (averaged over all relative angles between k and k') gives rise to a renormalization, so the effect survives even if the Coulomb interaction is screened. The enhancement follows a logarithmic dependence on carrier density, consistent with experimental data. Using an effective screening constant $\epsilon = \frac{\epsilon_{top}+\epsilon_{sub}}{2} = \frac{1+5}{2} = 3$, the theory predicts an averaged renormalization factor of 1.3-1.4 at carrier density $n = 1 - 2 \times 10^{12} cm^{-2}$ . Including intrinsic screening in graphene would slightly increase the effective $\epsilon$ to ~6, bringing the theory into better quantitative agreement with our measurements.

Fig. 4c presents the corresponding calculation for BLG, modeled within a two-band approximation with realistic hopping parameters. Two differences of physics arise: Unlike MLG, BLG has a nonzero density of states at the CNP, so the Thomas-Fermi screening length remains finite even at charge neutrality, giving a Coulomb interaction of the form $v_q = e^2/2\epsilon\epsilon_0(q + \eta q_{TF})$, where $\eta$ is a phenomenological parameter interpolating between the full static Thomas-Fermi limit ($\eta = 1$) and weaker screening. In the $\eta = 1$ case, the Coulomb interaction is short-ranged, and the Drude-weight renormalization is suppressed. Further, the different winding of wave functions means that only odd harmonics of the interaction contribute, and these are absent if the $k_F \ll q_{TF}$. The standard estimates would lead to a negligible effect, but our experiments show enhancements of 25-50% at carrier density $1 - 2 \times 10^{12} cm^{-2}$, indicating higher harmonic components of the interaction, perhaps arising from an effective screening length longer than expected from microscopic estimates. The plasmon renormalization in BLG may also depend on sensitively on non-parabolic band - structure details and dynamic screening conditions. A complete quantitative theory that incorporates these effects remains an open problem for future work.

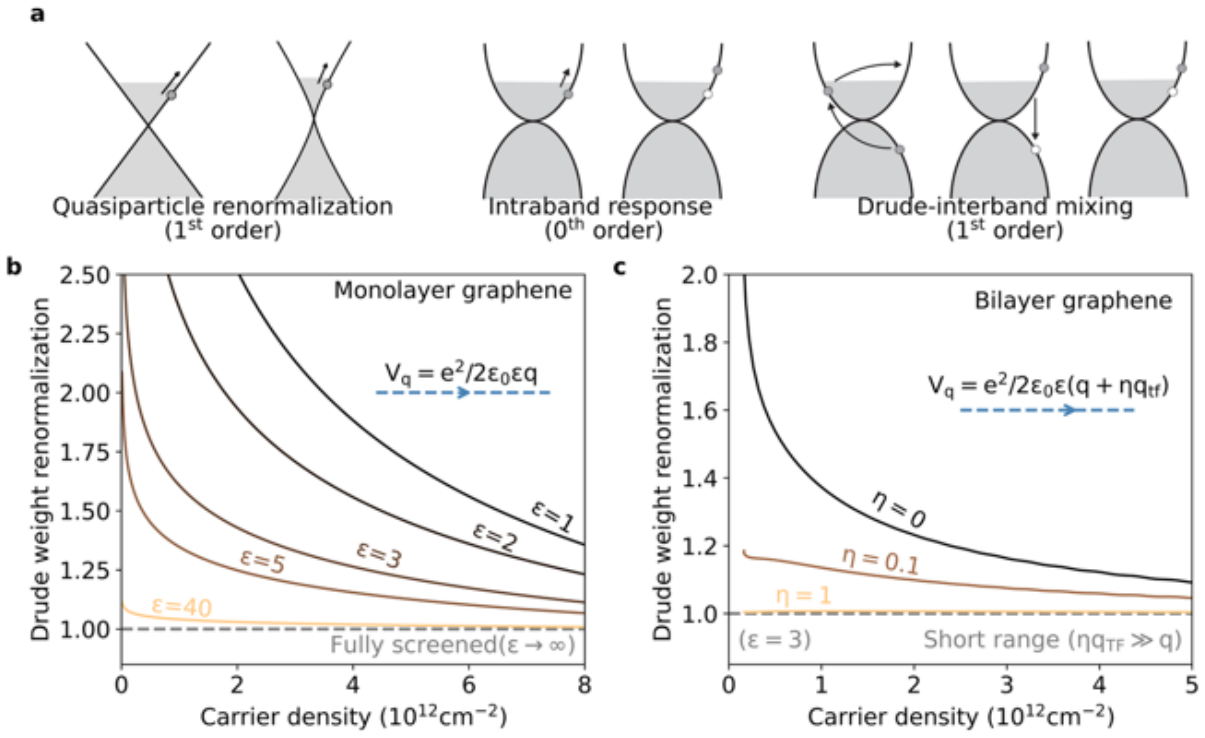

***Figure.** 4. Diagrammatic interpretation of Drude weight renormalization and first-order perturbation calculation. (a) Diagrammatic interpretation of many-body pathways for Drude weight enhancement. (b) Calculated Drude weight renormalization in monolayer. The false color scale on multiple curves represents long-range Coulomb interaction with varying dielectric constants $\epsilon$. (c) Calculated Drude weight renormalization in bilayer. The false color scale on the curve corresponds to a Thomas-Fermi approximation for different screening strength. A substrate screening constant $\epsilon = 3$ is used in the calculation.*

In this work, we demonstrate that the collective response in MLG and BLG differs markedly from conventional expectations, and we attributed the difference to the interplay of interactions and quantum geometry (wave function winding). An important outlook is to disentangle the interplay of electronic dispersion, interaction screening, and quantum geometry in collective charge dynamics of related systems.

Finally, we remark that the observed enhancement of Drude weight is a non-trivial phenomenon. Conventional many-body theory usually predicts that interaction suppress electronic motion, reducing the mobility, as commonly observed in ordinary solids. A notable recent counterexample is the enhancement of superfluid density in flat-band systems, which is attributed to the combined effects of interactions and nontrivial quantum geometry [36] . The Drude weight enhancement reported here represents a complementary manifestation. Extending space-time metrology to probing collective mode dynamics in other platforms, such as twisted bilayer graphene or moiré transition-metal dichalcogenides, are important directions for future investigations.

### ACKNOWLEDGMENTS

**Funding:**
Research on THz electrodynamics of graphene at Columbia is supported as part of Programmable Quantum Materials, an Energy Frontier Research Center funded by the U.S. Department of Energy (DOE), Office of Science, Basic Energy Sciences (BES), under award DE-SC0019443. Development of THz instrumentation at Columbia is supported by DOE-BES DE-SC0018426. Fabrication of plasmonic structures, is supported by W911NF2510062 (R.V., D.N.B.). The Flatiron institute is a division of the Simons foundation.

**Author contributions:**
S.X. and D.N.B. conceived the study. S.X. recorded the near-field data with support from R.A.V. . B.Y. prepared the graphene sample with guidance from C.D.. N.V. performed theoretical calculation with assistance from B.V., M.S.S., J.I., T.S., A.R., M.M.F, A.J.M, R.Q.. S.X. analyzed the data with assistance from N.V.,R.A.V,R.J.,Y.S.,M.D.,M.L.. S.X. and D.N.B. wrote the manuscript with input from all coauthors.

**Competing interests:**
The authors declare that they have no competing interests.

**Data and materials availability:**
All data needed to evaluate the conclusions in the paper are present in the paper and/or the Supplementary materials.